\documentstyle[12pt]{article}
\renewcommand{\baselinestretch}{1.15}
\begin{document}

\author{Mark Hindmarsh}
\date{May 1996}
\title{{\bf Analytic scaling solutions for cosmic domain walls}}
\begin{titlepage}

\maketitle

\vspace{-7cm}

\begin{flushright}
SUSX-TH-96-005\\
{\tt hep-ph/9605332}
\end{flushright}
\vspace{4cm}

\begin{center}

School of Mathematical and Physical Sciences\\
University of Sussex\\
Brighton BN1 9QH\\
U.K.\\
e-mail: {\tt m.b.hindmarsh@sussex.ac.uk}
\end{center}
\vfill

\begin{abstract}
A relativistic generalisation of a well-known method for approximating
the dynamics of topological defects in condensed matter is constructed,
and applied to the evolution of domain walls in a cosmological context.  
It is shown
that there are self-similar ``scaling'' solutions, for which one can in
principle calculate many quantities of interest without 
recourse to numerical simulations.  Here, the area density 
in the scaling regime is calculated in various backgrounds. 
Remarkably good agreement with numerical simulations 
is obtained.
\end{abstract}
\end{titlepage}

Topological defects formed at a cosmological phase transition are one of
the two known ways of creating scale-free primordial density
perturbations \cite{VilShe94,HinKib95}.  
However, unlike rival inflation-based
models \cite{KolTur90}, analytic calculations have not yet made much
impact.  There do exist analytic treatments of the defects and 
Goldstone modes arising from the spontaneous breaking
of global symmetries \cite{TurSpe91,FilBra94}, which are exact in the 
limit of large $N$, where
$N$ is the the number of scalar fields, but they remain relatively 
undeveloped.  

In this letter a new technique is outlined which promises to 
form the basis of
an exact dynamical theory for all
topological defects described by a Nambu-Goto action.  It is based on a
method well-known in the condensed matter literature: the $u$-theory of
Ohta, Jasnow and Kawasaki (OJK) \cite{OhtJasKaw82}, and 
its descendents \cite{DefScale}. In this approach,
the defects are replaced by a multicomponent scalar field (the ``$u$'' field) 
specially designed so that its
zeros track the positions of the defects.  Although the
equations of motion are non-linear, when combined with a Gaussian ansatz
for the field probability distribution, they are susceptible to a 
mean field theory treatment. One can then calculate analytically
many important quantities, such as the defect density and correlation
functions, purely from the two-point correlation function of the 
$u$-field, which is itself calculable.

To apply this theory to cosmic defects, all that is required is to make
it relativistically covariant.  The fictitious field that replaces the
defects still has a Gaussian ansatz for its probability distribution
function, and a self-consistent and self-similar solution for the
linearised equations of motion can be found.  With this in hand, one can
calculate the defect density, using a generalisation of
well-known techniques, although the presence of time derivatives of the 
$u$-field correlators complicates
the procedure somewhat.  In principle one could then
go on to calculate more or less anything of interest: for example, in
the cosmological context we would like to know the two-time correlation
functions of the cosmic string stress-energy tensor \cite{VeeSte90}. 
However, in this letter we shall
content ourselves with calculating the scaling value for the area
density of domain walls, in an approximation where one neglects 
the time derivatives of the $u$-field correlation function.  
The results for walls are 
encouraging when compared to the numerical simulations 
\cite{PreRydSpe89,CouLalOvr95,LarSarWhi96}. The results for 
strings are more difficult to obtain, and will be 
presented elsewhere \cite{Hin96}.

Another feature of the theory is that it describes the behaviour of
defects formed from initial conditions with a slight bias favouring 
one vacuum over another \cite{CouLalOvr95,LarSarWhi96}.  
It is found 
that the defects disappear at a conformal time $\eta_c = 
\eta_{\rm i} (U^2 / \langle u^2 \rangle_{\rm c})^{1/D}$ 
where $U=\langle u \rangle$ is the initial bias in the field, 
and $\langle u^2 \rangle_{\rm c}$ the initial fluctuations around that
value.  Indeed, part of the motivation for this work was to account for 
this behaviour observed in some interesting simulations recently 
carried out by Larsson, Sarkar and White \cite{LarSarWhi96}.

The essence of the technique is to replace the walls with a real scalar
field, which vanishes precisely at the coordinates of the walls
$X^\mu(\sigma^\alpha)$, where $\mu$ takes the values $0,\ldots,D$ and
$\alpha$ the values $0,\ldots,D-1$.  Thus we begin with the  equation
$u(X^\mu) = 0.$
Differentiating once with respect to the world-volume coordinates
$\sigma^\alpha$, we find
\begin{equation}
\partial_\beta X^\mu \partial_\mu u(X) = 0.
\label{eNorm}
\end{equation}
Thus the  vector $\partial_\mu u(X)$ is a spacelike normal
to the wall.  

The embedding of the $p$-brane in the background space-time induces a
metric in its world volume,
$
\gamma_{\alpha\beta} = g_{\mu\nu}(X)
\partial_\alpha X^\mu\partial_\beta X^\nu,
$
where $g_{\mu\nu}$ is the space-time metric. Using the embedding metric
we can covariantly differentiate (\ref{eNorm}) by acting with the operator
$(-\gamma)^{-1/2}\partial_\alpha(-\gamma)^{1/2}\gamma^{\alpha\beta}$, to
obtain
\begin{equation}
\Box X^\mu\partial_\mu u + \gamma^{\alpha\beta} 
\partial_\alpha X^\mu \partial_\beta X^\nu \partial_\mu\partial_\nu
u = 0,
\end{equation}
where $\Box$ is the covariant d'Alembertian and $\gamma = \det
\gamma_{\alpha\beta}$.  The defect equations of motion are
\cite{VilShe94,HinKib95}
\begin{equation}
\Box X^\mu + \Gamma^\mu_{\nu\rho} \gamma^{\alpha\beta} 
\partial_\alpha X^\mu \partial_\beta X^\nu = 0,
\label{eDEOM}
\end{equation}
where $\Gamma^\mu_{\nu\rho}$ is the affine connection. 
In equation (\ref{eDEOM}) we can identify the 
tensor $P^\parallel_{\mu\nu}=\gamma^{\alpha\beta}\partial_\alpha X_\mu
\partial_\beta X_\nu$, which is the tangential projector onto the 
wall.  We can replace this by $g^{\mu\nu}-P^\perp_{\mu\nu}$,
where 
$
P^\perp_{\mu\nu} = \partial_\mu u\partial_\nu u/(\partial u)^2.
$
Using the defect equations of motion (\ref{eDEOM}) 
we may then write
\begin{equation}
\left[(\partial u)^2 g^{\mu\nu} - \partial^\mu u\partial^\nu u\right] 
(\partial_\mu\partial_\nu u - \Gamma^\rho_{\mu\nu}\partial_\rho 
u) = 0.
\label{eFund}
\end{equation}
This is the fundamental equation of motion for the field
$u$ which replaces the defects.  It is actually unnecessary that
this field has anything to do with the underlying Higgs fields, and so
it may be called a ``fictitious'' field.  In other approaches, which
apply to global defects, the Higgs field $\phi$ is related to the fictitious
field by the non-linear transformation 
$\phi(\mbox{\boldmath $x$},t) = f(u)$, where
$f$ solves the static defect field equations as a function of 
the transverse coordinate $u$ \cite{DefScale}.

The equations of motion (\ref{eFund}) are not easy to solve, as they are
non-linear.  However, they have the distinct advantage over the original
equations of motion (\ref{eDEOM}) in that they are local:  the defects
may self-intersect and reconnect, a process which is highly non-local in
its world volume.  The approach taken in the condensed matter context is
essentially that of mean field theory: bilinears in the field $u$
are replaced by their averages.   Thus one is assuming that the field is
a Gaussian random field, and remains so throughout it evolution.  This
seems a good starting point for the relativistic version as well,
although it should be borne in mind that the approximation is not 
well-controlled.

We begin the mean field theory manipulations by defining the basic 
equal-time two-point correlation function:
\begin{equation}
\langle u(\mbox{\boldmath $x$},\eta)
u(\mbox{\boldmath $x$}',\eta)\rangle =  C(|\mbox{\boldmath $x$} -
\mbox{\boldmath $x$}'|,\eta),
\end{equation}
where the angle brackets denote an average over a spatially isotropic
Gaussian probability distribution function.  We shall also define
$M_{\mu\nu}$ to be the equal-time two-point correlator of $\partial_\mu u$:
\begin{equation}
\langle\partial_\mu u(\mbox{\boldmath $x$},\eta)
\partial_\nu u(\mbox{\boldmath $x$}',\eta)\rangle =  
M_{\mu\nu}(|\mbox{\boldmath $x$}-\mbox{\boldmath $x$}'|,\eta).
\end{equation}
A two-point  correlator with three derivatives will also be useful:
\begin{equation}
\langle \partial_\mu u(\mbox{\boldmath $x$},\eta)
\partial_\nu\partial_\rho u(\mbox{\boldmath $x$}',\eta)\rangle = 
\gamma_{\mu\nu\rho}(|\mbox{\boldmath $x$}-\mbox{\boldmath $x$}'|,\eta).
\end{equation}
When referred to with no explicit spatial variables, the 
correlators are to be taken at zero separation. In this coincident 
limit, the assumed spatial isotropy of the distribution function dictates 
their forms. The non-zero components of $M_{\mu\nu}$ are 
\begin{equation}
M_{00} = T(\eta), \qquad M_{mn} = S(\eta)\delta_{mn}, 
\end{equation}
where $m,n = 1,\ldots,
D$, and the non-zero components of $\gamma_{\mu\nu\rho}$ are 
\begin{equation}
\gamma_{000}(\eta) =  \frac{1}{2}\dot T(\eta), \qquad
\gamma_{0mn}(\eta)  =  -\frac{1}{2}\dot S(\eta)\delta_{mn}, \qquad
\gamma_{m0n}(\eta)  =  \frac{1}{2}\dot S(\eta)\delta_{mn}.
\end{equation}
We now linearise the equations of motion by 
taking the Gaussian average, and then find a self-consistent 
solution for the fields $u(\mbox{\boldmath $x$},\eta)$.  We need the
following identities:
\begin{eqnarray}
\langle (\partial u)^2 \, \partial_\mu\partial_\nu u\rangle & = & 
M \partial_\mu\partial_\nu u +2
\gamma_{\rho\mu\nu} g^{\rho\sigma} \partial_\sigma u
\label{eExp1}\\
\langle (\partial u)^2\, \partial_\rho u\rangle & = & M
\partial_\rho u + 2 g^{\rho\sigma}M_{\eta\rho} \partial_\sigma u,
\label{eExp2}
\end{eqnarray}
where $M=M_{\mu\nu}g^{\mu\nu}$.

In a flat Friedmann-Robertson-Walker space-time, the 
affine connection is 
$
\Gamma^\rho_{\mu\nu} = (\delta_\mu^\rho\delta_\nu^0 
+ \delta_\nu^\rho\delta_\mu^0 - g_{\mu\nu}g^{\rho 0})(\dot a/a)$.
We can now see that the linearised equations of motion have the 
form
\begin{equation}
\ddot u + \frac{\mu(\eta)}{\eta} \dot u - v^2\nabla^2u = 0,
\label{eLEOM}
\end{equation}
where $\mu(\eta)$ and $v$ depend on $D$.
For Friedmann models, one can show that 
\begin{eqnarray}
\mu(\eta) &= &	-2\eta(\dot S /S) + 
	{\alpha(\eta)}D\left[ 1 - 
	3\left({T}/{S}\right)\right],
\\
v^2 &=& 	\left[{D-1} - \left({T}/{S}\right)\right]/D,
\end{eqnarray}
where $\alpha(\eta) = \eta\dot a/a$.

In a scaling solution we would expect $S$ and $T$ to have 
a power-law behaviour with $\eta$.  Thus, 
so as long as we are not near a transition in the equation of state 
of the Universe, $\mu$ and $v^2$ are constant. Imposing the boundary 
condition that $u$ be regular as $\eta \to 0$,  
(\ref{eLEOM}) then has the simple solution
\begin{equation}
u_{\mbox{\boldmath $k$}}(\eta)  = A_\nu 
\left(\frac{\eta}{\eta_{\rm i}}\right)^{(1-
\mu)/2+\nu} \frac{J_\nu(kv\eta)}{(kv\eta)^\nu},
\end{equation}
where $A_\nu = 2^\nu \Gamma(\nu+1) u_{\mbox{\boldmath $k$}}(\eta_{\rm i})$, 
and $(1-\mu)^2/4 = \nu^2$. The form of the initial power spectrum 
$P_{\rm i}(k) = |u_{\mbox{\boldmath $k$}}(\eta_{\rm i})|^2$ is taken to be 
white noise, which ensures that the field is spatially uncorrelated 
to begin with.

We may now evaluate $T/S$ and $v^2$, and implicitly 
solve for $\nu$.  Firstly, we must decide the sign of $\nu$. It 
turns out that it is inconsistent to take $(1-\mu)/2 = \nu$ (one 
way of showing this is to compare $\dot C(\eta)$ calculated 
by explicit differentiation, and by evaluation of $\langle u 
\dot u\rangle$).  Thus $C$ scales as $\eta^{-D}$, $S$ and $T$ as 
$\eta^{-(D+2)}$.
Using standard integrals of Bessel functions, 
and defining the parameter $\beta = 2\nu - D - 1$, we find 
\begin{equation}
\frac{T}{S} =  \frac{(D+2)(D-2)}{3(D+2)+2\beta}, 
\end{equation}
provided $\beta > 0$, so that the integrals for $S$ and $T$ 
are defined. To find $\beta$, we solve the equation
\begin{equation}
\mu \equiv \beta+D+2 = 
	\alpha D\left[1 - 3(T/S)\right] + 
	2(D+2).
\end{equation}
The radiation-dominated 
$(\alpha =1)$ and
matter-dominated $(\alpha=2)$ Friedmann models require a certain 
amount of algebra: however, Minkowski space ($\alpha=0$) has the simple 
solution $\beta=D+2$, giving $T/S=(D-2)/5$ and $v^2 = (4D-3)/5D$. 
The smallness of $T/S$ will help us in the next part of the
argument, where neglecting $T$ (and another 
correlator involving time derivatives) in comparison to
$S$ will make calculations much easier.

Armed with the ``mean-field'' solution for 
$u(\mbox{\boldmath $x$},\eta)$ we can now
calculate anything that can be expressed in terms of local functions of 
the field and its derivatives, provided of course that we are able to
perform the Gaussian integrals involved.  In this letter we content
ourselves merely with evaluating the area density, which is given by
\begin{equation}
{\cal A} = \int d^D\sigma\,\sqrt{-\gamma}\, \delta_{D+1} 
(x-X(\sigma)).
\end{equation}
Making the coordinate transformation from $x^\mu$ to $(\sigma^\alpha,
u)$ near the wall, this can be rewritten as 
\begin{equation}
{\cal A}  = \delta(u)|\partial u|.
\end{equation}
In order to calculate the Gaussian average of ${\cal A} $, 
one takes Fourier transforms \cite{OhtJasKaw82}:
\begin{equation}
\langle {\cal A} \rangle  = 
 - \frac{\Gamma(D/2)}{2\pi^{1+D/2}}
\int \frac{dk}{2\pi} \int \frac{d^{D+1}q}{(q^2)^{D/2}}\,
\left\langle
\frac{\partial}{\partial q}\cdot\frac{\partial}{\partial q}
\, e^{iq\cdot\partial u+iku}  \right\rangle.
\label{eDensity}
\end{equation}
One potential difficulty is that $q^2$ can vanish, due to the Lorentzian 
metric. The equation must therefore formally be defined 
by analytic continuation to imaginary time, and an accompanying 
Wick rotation of the time components of the Fourier transform 
variable $q^\mu$.

We neglect terms of order $T/S$, which complicate the calculation 
considerably. With less justification, 
we shall also neglect terms
involving ${\dot C}^2/CS$ which appear in the full calculation, and 
are of order 0.5, as one can check from the formula (\ref{eSoC})
for $S/C$ below.  The averaged comoving defect area density 
${\cal A}(\eta)$ is then just 
\begin{equation}
\langle {\cal A}(\eta) \rangle = \sqrt{2}\,\frac{\Gamma((D+1)/2)}
{\Gamma(D/2)} \left(\frac{S}{2\pi C}\right)^{1/2} + 
O(T/S,\dot C^2/CS).
\label{eArea}
\end{equation}
This is identical to condensed matter results for domain walls 
\cite{OhtJasKaw82,ToyHon86,LiuMaz92}.

One can show that 
\begin{equation}
\frac{S}{C} = \frac{1}{\eta^2}\frac{(\beta+1)(\beta+1+D/2)}{2\beta v^2},
\label{eSoC}
\end{equation}
and we immediately see that the correct scaling 
behaviour for the wall area density (${\cal A}(\eta) \propto \eta^{-1}$) 
 is reproduced.  Furthermore, we are 
able to compute the coefficient, and compare with numerical 
simulations.  The comoving area densities for walls 
in two- and three-dimensional Minkowski space, radiation- and matter-dominated 
Friedmann models
are displayed in  Table \ref{tValues}.  

\begin{table}
\begin{center}
\begin{tabular}{|c|c|c|}
\hline
$\alpha$ & ${\cal A}$ ($D=2$)&  ${\cal A}$ ($D=3$)\\
\hline
0 &   $1.4 \eta^{-1}$   & $1.7 \eta^{-1}$ \\
1 &   $1.5 \eta^{-1}$   & $1.8 \eta^{-1}$ \\
2 &   $1.7 \eta^{-1}$   & $2.0 \eta^{-1}$ \\
\hline
\end{tabular}
\caption{\small \setlength{\baselineskip}{12pt}
The calculated values of the comoving 
domain wall area density in the self-consistent 
linearised solution for $u$.
Values listed 
are for Minkowksi space $(\alpha=0)$, and 
radiation and matter-dominated FRW models ($\alpha=1,2$
respectively).
\label{tValues}}
\end{center}\end{table}

The theory seems to work  well:
scalar field theory simulations in $D=3$ give a comoving 
area density of approximately $1.5/\eta$ in both radiation 
\cite{CouLalOvr95,LarSarWhi96} and 
matter \cite{PreRydSpe89} eras, which 
sits nicely with the calculated values.  However, 
the agreement is partly fortuitous, as there are probably significant  
errors in both the theoretical and numerical values.  In $D=2$ the agreement 
is also good, although 
there is evidence for a small deviation from the $\eta^{-1}$ 
behaviour in $D=2$ \cite{PreRydSpe89,CouLalOvr95}.

One may also ask how the network behaves when a small 
bias is introduced into the initial conditions, that is, if 
$\langle u(\mbox{\boldmath $x$},\eta_{\rm i}) \rangle = U$.  
In numerical simulations of domain 
walls \cite{CouLalOvr95,LarSarWhi96}, it is found that 
even for very small initial 
biases, for which the walls percolate, the system still evolves 
away from the percolating state and eventually the large walls 
break up and disappear. 
Similar behaviour is well-known in 
in the study of quenches of condensed matter systems with a
non-conserved order parameter 
\cite{OhtJasKaw82,ToyHon86,MonGol92,FilBraPur95}. 

The theoretical description of this behaviour is fairly straightforward.
Introducing a bias into the initial conditions alters the Gaussian
average of (\ref{eDensity}) to
$
\langle {\cal A}(\eta) \rangle_U= \langle {\cal A}(\eta) \rangle \exp\left(
-U^2/2C(\eta)\right)
$,
where $\langle {\cal A}(\eta) \rangle$ is the zero bias result from 
(\ref{eArea}).
If the system is close to being self-similar at some initial time
$\eta_{\rm i}$ when the magnitude of the bias is $U$ and the fluctuation
around that value is $C(\eta_{\rm i})$, one can calculate the time 
$\eta_{\rm c}$ at which the defect density falls to a fraction $e^{-1}$
of its scaling value to be 
\begin{equation}
\eta_{\rm c} = \eta_{\rm i}\left(U^2/2C(\eta_{\rm i})\right)^{-1/D}.
\label{eDisTime}
\end{equation}
Recent simulations by Larsson, Sarkar and White are consistent with 
the above calculations of $\langle {\cal A}(\eta) \rangle_U$ and 
$\eta_{\rm c}$ in $D=2$, but do not have 
sufficiently good statistics in $D=3$ to be able 
to check the results \cite{LarSarWhi96}.  Coulson 
et al.~\cite{CouLalOvr95} did not attempt fits of the correct form, ${\cal A} 
\propto \eta^{-1}\exp(-(\eta/\eta_{\rm c})^D)$, 
to their simulations, although they did note that the walls disappeared faster 
than a simple exponential in $D=3$.

To summarise, this paper outlines a new analytic technique for describing the 
dynamics of topological defects after a cosmological phase transition.
It is a relativistic version of a well-known approach in condensed
matter physics, due to Ohta, Jasnow and Kawasaki \cite{OhtJasKaw82}. The
scaling area density of domain walls in three spatial dimensions is 
calculated, and agrees quantitatively with numerical simulations of a 
real scalar field \cite{PreRydSpe89,CouLalOvr95,LarSarWhi96}.  
Further work is clearly needed, particularly on strings, 
but it already seems clear that developments 
of this technique will be of great benefit to the study of cosmic
defects.

I am deeply indebted to Sebastian Larsson, Subir Sarkar, and Peter White for
communicating their results to me in advance of publication, and for
useful discussions.  The seeds of this work lie in the Topological 
Defects programme at the Isaac Newton Institute, Cambridge, July--Dec 
1994.  I am also grateful for the hospitality of the NORDITA/Uppsala 
Astro-Particle Workshop 1996, where this work was completed.
The author is supported by PPARC Advanced
Fellowship number B/93/AF/1642, by PPARC grant GR/K55967, and by 
the European Commission under the Human Capital and Mobility
programme, contract no.~CHRX-CT94-0423.

\newpage
\renewcommand{\baselinestretch}{1}

\end{document}